\documentclass[12pt, a4paper]{article}

\usepackage[pdftex]{graphicx}
\usepackage{bm,afterpage,amsmath,placeins,flafter,authblk,adjustbox,lscape,caption}
\usepackage[english]{babel}
\usepackage{setspace}
\usepackage{booktabs}
\usepackage{changes}
\usepackage[top=1.3in,left=1.1in,right=1.1in,bottom=1.3in]{geometry}
\usepackage{color}

\graphicspath{{figures/}}

\date{}

\begin{document}
	
	\title{Space-Time Analysis of Movements in Basketball using Sensor Data}
	
	\author[1]{Rodolfo Metulini}
	\author[1]{Marica Manisera}
	\author[1]{Paola Zuccolotto}

	\affil[1]{\small Department of Economics and Management, University of Brescia}

\maketitle

\textbf{Please cite as:} \textit{Metulini, R., Manisera, M., Zuccolotto, P. (2017), Space-Time Analysis of Movements in Basketball using Sensor Data, "Statistics and Data Science: new challenges, new generations" SIS2017 proceeding. Firenze Uiversity Press. e-ISBN: 978-88-6453-521-0}

\begin{abstract}
Global Positioning Systems (GPS) are nowadays intensively used in Sport Science as they permit to capture the space-time trajectories of players, with the aim to infer useful information to coaches in addition to traditional statistics. In our application to basketball, we used Cluster Analysis in order to split the match in a number of separate time-periods, each identifying homogeneous spatial relations among players in the court. Results allowed us to identify differences in spacing among players, distinguish defensive or offensive actions, analyze transition probabilities from a certain group to another one.
\end{abstract}


\section{Introduction}
\label{sec:intro}

Studying the interaction between players in the court, in relation to team performance, is one of the most important issue in Sport Science. In recent years, thanks to the advent of Information Technology Systems (ITS), it became possible to collect a large amount of different types of spatio-temporal data, which are, basically, of two kinds. On the one hand, play-by-play data report a sequence of relevant events that occur during a match. Events can be broadly categorized as player events such as passes and shots; and technical events, for example fouls and time-outs. Carpita et al. \cite{carpita2013,carpita2015} used cluster analysis and principal component analysis in order to identify the drivers that affect the probability to win a football match. Social network analysis has also been used to capture the interactions between players \cite{wasserman1994social}; Passos et al. \cite{passos2011networks} used centrality measures with the aim of identifying central players in water polo. On the other hand, object trajectories capture the movement of players or the ball. Trajectories are captured using optical- or device-tracking and processing systems. Optical systems use cameras, the images are then processed to compute the trajectories \cite{bradley2007reliability}, and commercially supplied to professional  teams or leagues \cite{Tracab,Impire}. Device systems rely on devices that infer location based on Global Positioning Systems (GPS) and are attached to the players' clothing \cite{Catapult}. The adoption of this technology and the availability of data is driven by various factors, particularly commercial and technical. Even once trajectories data become available, explaining movement patterns remains a complex task, as the trajectory of a single player depends on a large amount of factors. The trajectory of a player depends on the trajectories of all other players in the court, both teammates and rivals. Because of these interdependencies a player action causes a reaction.  A promising niche of Sport Science literature, borrowing from the concept of Physical Psychology \cite{turvey1995toward}, expresses players in the court as agents that face with external factors \cite{travassos2013performance,araujo2016team}. In addition, typically, there are certain role definitions in a sports team that influence movement. Predefined plays are used in many team sports to achieve specific objectives; moreover, teammates who are familiar with each other's playing style may develop productive interactions that are used repeatedly. Experts want to explain why, when and how specific movement behavior is expressed because of tactical behavior and to retrieve explanations of observed cooperative movement patterns. A common method to approach with this complexity in team sport analysis consists on segmenting a match into phases, as it facilitates the retrieval of significant moments of the game. For example, Perin et al. \cite{perin2013soccerstories} developed a system for visual exploration of phases in football. Using a basketball case study, and having available the spatio-temporal trajectories extracted from GPS tracking systems, this paper aims at characterizing the spatial pattern of the players in the court by defining different game phases by mean of a cluster analysis. We identify different game phases, each considering moments being homogenous in terms of spacings among players. First, we characterize each cluster in terms of players' position in the court. Then, we define whether each cluster corresponds to defensive or offensive actions and compute the transition matrices in order to examine the probability of switching to another group from time $t$ to time $t+1$.
 
\section{Data and Methods}
\label{sec:meth}

Basketball is a sport generally played by two teams of five players each on a rectangular court ($28m x 15m$). The match, according to International Basketball Federation (FIBA) rules, lasts $40$ minutes, and is divided in four periods of $10$ minutes each. The objective is to shoot a ball through a hoop $46 cm$ in diameter and mounted at a height of $3.05m$ to backboards at each end of the court.  The data we used in the analyses that follow refers to a friendly match played on March 22th, 2016 by a team based in the city of Pavia (Italy). This team played the 2015-2016 season in the C-gold league, the fourth league in Italy. Totally, six players took part to the friendly match. All those players worn a microchip in their clothings. The microchip collects the position (in pixels of 1 $m^2$) in both the $x$-axis and the $y$-axis, as well as in the $z$-axis (i.e. how much the player jumps). The positioning of the players has been detected at millisecond level. Considering all the six players, the system recorded a total of $133,662$ space-time observations ordered in time. In average, the system collects positions about $37$ times every second. Considering that six players are in the court at the same time, the position of each single player is collected, in average, every $162$ milliseconds.  $x$-axis (length) and $y$-axis (width) coordinates have been filtered with a Kalman approach. The Kalman filtering is an algorithm used to predict the future state of a system based on the previous ones, in order to produce more precise estimates. We cleaned the dataset by dropping the pre-match, the half-time break and the post match periods. Then, the dataset has been modified by adding all the milliseconds that have not been detected. We attributed to these moments the value of the coordinates of the first previous instant with non-missing values. We then reshaped the dataset in order that the dataset rows are uniquely identified by the millisecond, and each player's variable is displayed in a specific column. The final dataset counts for $3,485,147$ total rows. We applied a $k$-means Cluster Analysis in order to group a set of objects. Cluster analysis is a method of grouping a set of objects in such a way the objects in the same group (clusters) are more similar to each other than to those in other groups. In our case, the objects are represented by the time instants, expressed in milliseconds, while the similarity is expressed in terms of players' distance\footnote{In the analyses that follows, for the sake of simplicity, we only consider the period where player 3 was in the bench.}. Based on the value of the between
deviance (BD) / total deviance (TD) ratio and the increments of this value
by increasing the number of clusters by one, we chose $k$=8 (BD/TD=50\% and relatively low increments for increasing $k$, for $k$$\ge$8)

\section{Results}
\label{sec:res}
\begin{figure}[!htb]
	\centering
	\includegraphics[scale=0.6]{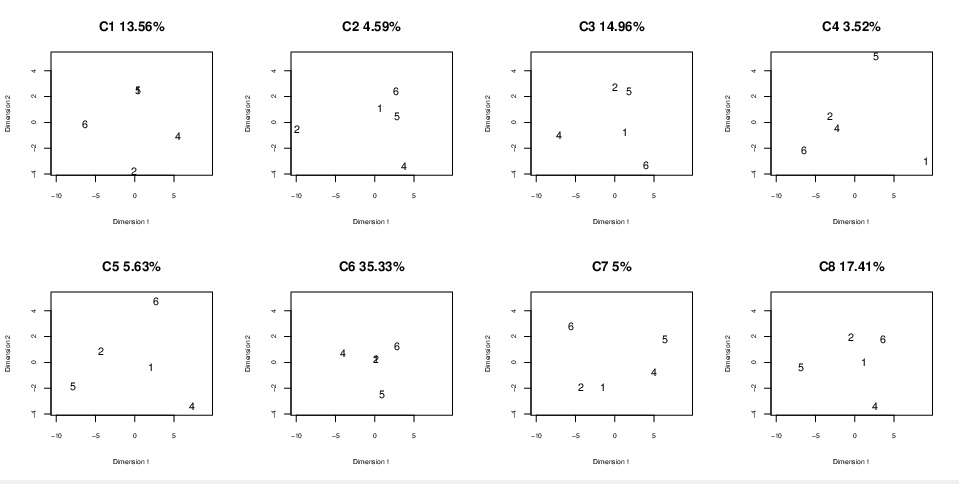}
	\caption{Map representing, for each of the 8 clusters, the average position in the $x-y$ axes of the five players, using MDS.}
	\label{fig:mds}
\end{figure}

The first cluster (C1) embeds 13.56\% of the observations (i.e. 13.56\% of the total game time). The other clusters, named C2, ..., C8, have size of 4.59\%, 14.96\%, 3.52\%, 5.63\%, 35.33\%, 5.00\% and 17.41\% of the total sample size, respectively. We used Multidimensional Scaling (MDS) in order to plot the differences between the groups in terms of positioning in the court. With MDS algorithm we aim to place each player in $N$-dimensional space such that the between-player average distances are preserved as well as possible. Each player is then assigned coordinates in each of the $N$ dimensions. We choose $N$=2 and we draw the related scatterplots as in Figure \ref{fig:mds}. We observe strong differences between the positioning pattern among groups. In C1 and C5 players are equally spaced along the court. C6 also highlights an equally spaced structure, but the five players are more closed by. In other clusters we can see a spatial concentration: for example in C2 players 1, 5 and 6 are closed by while in C8 this is the case of players 1, 2 and 6.  

\begin{figure}[!htb]
	\centering
	\includegraphics[scale=0.6]{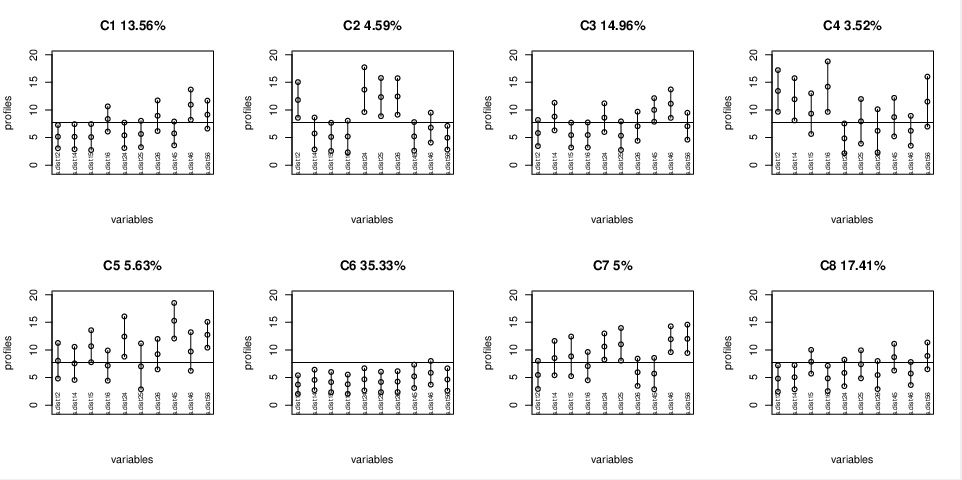}
	\caption{Profile plots representing, for each of the 8 clusters, the average distance among each pair of players.}
	\label{fig:profplot}
\end{figure}

\begin{figure}[!htb]
	\centering
	\includegraphics[scale=.8]{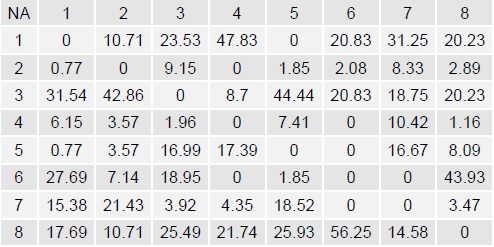}
	\caption{Transition matrix reporting the relative frequency subsequent moments  ($t$, $t + 1$) report a switch from a group to a different one.}
	\label{fig:transmat_ad}
\end{figure}

Figure \ref{fig:profplot} reports cluster profile plots and helps us to better interpret the spacing structure in Figure \ref{fig:mds}, characterizing groups in terms of average distances among players. Profile plot for C6 confirms that players are more close by, in fact, all the distances are smaller than the average distance. At the same way, C2 presents distances among players 1, 5 and 6 smaller than the average. 
After having defined whether each moment corresponds to an offensive or a defensive action looking to the average coordinate of the five players in the court, we also found that some clusters represent offensive actions rather than defensive. More precisely, we found that clusters C1, C2, C3 and C4 mainly correspond to offensive actions (respectively, for the 85.88\%, 85.91\%, 73.93\% and 84.62\% of the times in each cluster) and C6 strongly corresponds to defensive actions (85.07\%). Figure \ref{fig:transmat_ad} shows the transition matrix, which reports the relative frequency in which subsequent moments in time report a switch from a cluster to a different one. It emerges that for the 31,54\% of the times C1 switches to a new cluster, it switches to C3, another offensive cluster. C2 switches to C3 for the 42.85\% of the times. When the defensive cluster (C6) switches to a new cluster, it switches to C8 for the 56.25\% of times. 

\section{Conclusions and future research}
\label{sec:conc}
In recent years, the availability of `big data" in Sport Science increased the possibility to extract insights from the games that are useful for  coaches, as they are interested to improve their team's performances. In particular, with the advent of Information Technology Systems, the availability of players' trajectories permits to analyze the space-time patterns with a variety of approaches: Metulini \cite{metulini2016motion}, for example, adopted motion charts as a visual tool in order to facilitate interpretation of results. Among the existing variety of methods, in this paper we used a cluster analysis approach based on trajectories' data in order to identify specific pattern of movements. We segmented the game into phases of play and we characterized each phase in terms of spacing structure among players, relative distances and whether they represent an offensive or a defensive action, finding substantial differences among different phases.These results shed light on the potentiality of data-mining methods for trajectories analysis in team sports, so in future research we aim to i) extend the analysis to multiple matches, ii) match the play-by-play data with trajectories in order to extract insights on the relationship between particular spatial patterns and the team's performances.

\section*{Acknowledgments}
Research carried out in collaboration with the Big\&Open Data Innovation Laboratory (BODaI-Lab), University of Brescia (project nr. 03-2016, title Big Data Analytics in Sports, www.bodai.unibs.it/BDSports/), granted by Fondazione Cariplo and Regione Lombardia. Authors would like to thank MYagonism (https://www.myagonism.com/) for having provided the data.

%
%
%

\end{document}